\documentclass[twocolumn,aps,superscriptaddress,showpacs,preprintnumbers,amsmath,amssymb]{revtex4}
\usepackage{graphicx} 
\usepackage{bm} 

\newcommand{\bra}[1]{\langle #1 |}
\newcommand{\ket}[1]{| #1 \rangle}

\newcommand{\beq}{\begin{equation}}
\newcommand{\eeq}{\end{equation}}
\newcommand{\crea}[3]{\hat #1^{\dagger}_{#2,#3}}
\newcommand{\anni}[3]{\hat #1_{#2,#3}}

\newcommand{\creation}[2]{\hat #1^{\dagger}_{#2}}
\newcommand{\annihilation}[2]{\hat #1_{#2}}

\def\vec#1{\mathbf{#1}}
\def\k{\mathbf{k}}
\def\q{\mathbf{q}}
\def\xi{\mathbf{x}_i}

\def\pra#1#2#3{Phys.~Rev.~A~{\bf #1},\ #2\ (#3)}
\def\prl#1#2#3{Phys.~Rev.~Lett.~{\bf #1},\ #2\ (#3)}

\begin{document}

\title{Tunable disorder in a crystal of cold polar molecules}

\author{Felipe Herrera}
\affiliation{Department of Chemistry, University of British Columbia, Vancouver, B.C., V6T 1Z1, Canada}

\author{Marina Litinskaya}
\affiliation{Department of Chemistry, University of British Columbia, Vancouver, B.C., V6T 1Z1, Canada}
\affiliation{ Institute of Spectroscopy RAS, Troitsk, Moscow Region 142190, Russia}

\author{Roman V. Krems}
\affiliation{Department of Chemistry, University of British Columbia, Vancouver, B.C., V6T 1Z1, Canada}
\date{\today}

\maketitle


There is a growing interest in designing tunable many-body quantum systems, which can be used as physical simulators to access novel phenomena and to study the properties of analagous, but less controllable, many-body quantum systems \cite{Lewenstein:2007, Micheli:2006,greiner,Gorshkov:2010}. For example, a major thrust of current research is focussed on realizing lattice-spin models with ultracold molecules trapped on optical lattices \cite{Micheli:2006,greiner,Rabl:2007,Pupillo:2008,Pollet:2010}. 
This quantum simulation work relies on tuning interactions of ultracold molecules by external fields.
While two-body interactions of ultracold molecules can be tuned by a variety of methods \cite{pccp}, it is much harder to achieve external field control over many-body interactions.  
In the present work, we demonstrate the possibility of controlling by an external field the dynamics of collective excitations (excitons) of molecules on an optical lattice. We show that a suitably chosen two-species mixture of ultracold polar molecules loaded on an optical lattice forms a phononless crystal,  where exciton-impurity interactions can be controlled by applying an external electric field. 
This can be used for the controlled creation of many-body entangled states of ultracold molecules and  the time-domain quantum simulation of disorder-induced localization and delocalization of quantum particles.


The present work builds on three major recent advances in atomic, molecular and optical physics: the creation of dense ensembles of ultracold polar molecules in the ro-vibrational ground state by photoassociation of alkali metal atoms \cite{Ni:2008, Deiglmayr:2008}; the development of techniques for trapping ultracold atoms and molecules by optical lattices \cite{Bloch:2005}; and the development of techniques for detecting and manipulating single atoms in optical lattices \cite{Wurtz:2009, Karski:2009, Bakr:2009}. It has thus become technologically possible to create ordered ensembles of ultracold polar molecules in the ro-vibrational ground state trapped by an optical lattice with a lattice separation of about 400 nm \cite{Danzl:2010}. Using the method of Ref. \cite{Wurtz:2009}, molecules in specific lattice sites can be evaporated. By overlapping two optical lattices of trapped molecules or using the method of Ref. \cite{Rabl:2003}, it should be possible to create a mixture of molecules arranged in an arbitrary array with any spatial dimensionality.

\begin{figure}[t]
\includegraphics[width=0.47\textwidth]{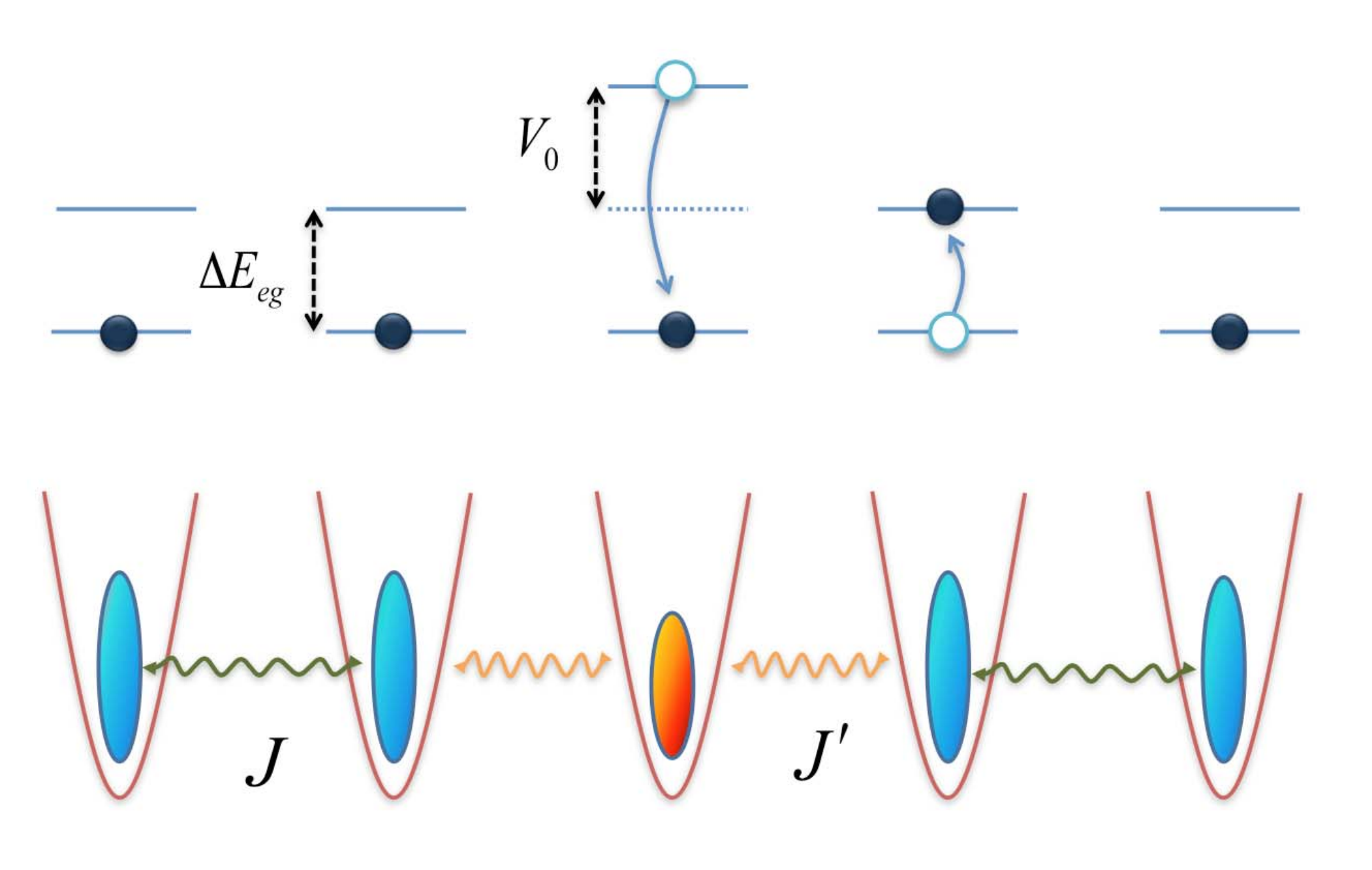}
\caption{Schematic illustration of a crystal with tunable impurities.  
Molecules are confined by the periodic potential of an optical lattice and coupled by the dipole-dipole interaction, which allows the exchange of a rotational excitation between molecules in different lattice sites. An impurity is a molecule with a different dipole moment and different rotational energy splitting. For certain combinations of molecules, the energy defect $V_0$ can be tuned by an external electric field from zero to a large magnitude, thereby modifying the energy transfer in the crystal.  
}
\label{fig:dispersion curves}
\end{figure}

Figure 1 illustrates the system considered here. We consider an optical lattice of $^1\Sigma$ polar molecules, such as LiCs, with one molecule in the rovibrational ground state per lattice site. We assume that the tunneling of molecules to different lattice sites is entirely suppressed, which can be easily achieved by applying laser fields of sufficient power \cite{Bloch:2005}.  The lattice sites are separated by 400 nm so the interaction between the molecules is entirely determined by the long-range dipole-dipole interaction potential. 
We consider the lowest excited states of the molecular crystal. In the absence of the dipole - dipole interactions, these would be the rotational excitations of the individual molecules. However, the interactions between the molecules lead to collective modes known as Frenkel excitons \cite{Agranovich:2008}. Each exciton state is a plane wave characterized by a wavevector.  
In the absence of an electric field, the rotational ground state of a $^1\Sigma$ diatomic molecule is characterized by the rotational angular momentum $N=0$. The first excited rotational state $N=1$ is triply degenerate. The $N=0 \rightarrow N=1$ excitation of molecules therefore leads to three excitonic states characterized by the dispersion curves shown in Figure 2. In the presence of an electric field, the rotational state $N=1$ splits into two Stark energy levels corresponding to the projections $M=0$ and $M=\pm 1$ of $\hat{N}$ on the electric field axis. This lifts the degeneracy of the exciton states as shown in Figure 2a. When the electric field is perpendicular to the molecular array, excitons ${\gamma}$ and ${\beta}$ have negative effective mass, i.e., their energy decreases with increasing wavevector, while the effective mass of exciton $ \alpha $ is positive. The effective mass can be changed by varying the strength and the direction of the external electric field (see Figure 2b).  We also note that, unlike electronic excitons in molecular crystals, rotational excitons have a very long radiative lifetime exceeding seconds. In the following, we focus on the exciton state labeled by $\gamma$ in Figure 2a.

\begin{figure}[t]
\includegraphics[width=0.47\textwidth]{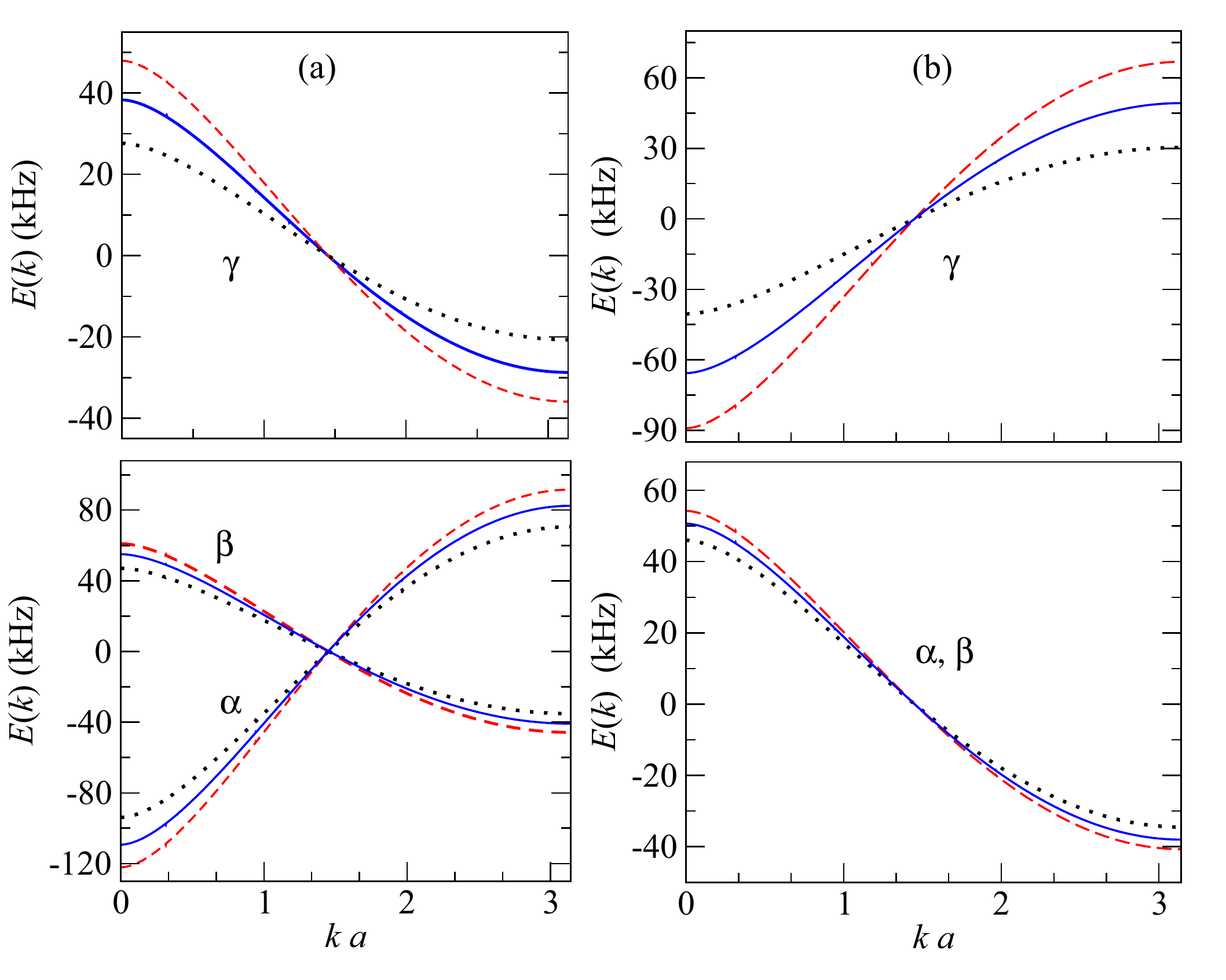}
\caption{Stark effect on rotational excitons. 
Exciton energy $E$($k$) calculated for three lowest excitations in a 1D array of $10^5$ LiCs molecules separated by 400 nm as functions of the exciton wavenumber ${k}$. Exciton states $\alpha$, $\beta$ and $\gamma$ centered at the isolated molecule transition energies $\Delta E_{eg}$ are shown in the presence of an electric field perpendicular to the array (panel a) and parallel to the array (panel b). The electric field magnitudes are 2 kV/cm (dotted lines), 3.2 kV/cm (solid lines), 5 kV/cm (dashed lines).  At these electric fields, exciton $\gamma$ is separated from excitons $\alpha$ and $\beta$ by the energy much larger than the exciton bandwidths. The figure shows that the magnitude and the sign of the effective mass of the rotational excitons can be controlled by varying the magnitude and the direction of the electric field.
}
\label{fig:dispersion curves}
\end{figure}
\begin{figure}[ht]
\includegraphics[width=0.47\textwidth]{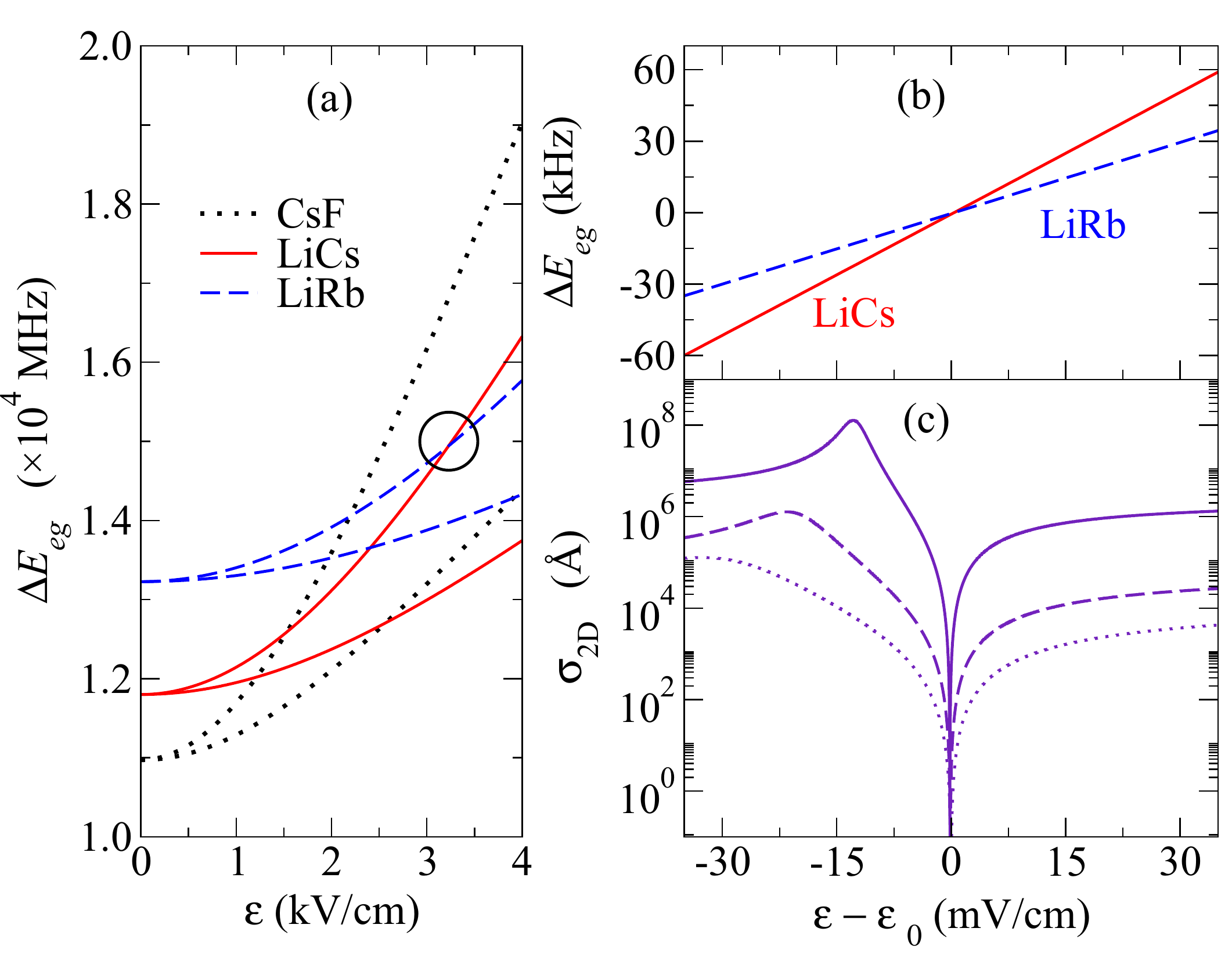}
\caption{
Excitation energies of non-inetracting molecules in an electric field. Panel (a) shows  the excitation energies $\Delta E_{eg}$ for transitions $\ket{N = 0, M= 0}\rightarrow\ket{N= 1,M}$ with $M = 0$ (upper curve) and $M=\pm 1$ (lower curve) vs electric field for three polar molecules. Panel (b) shows an expanded view of the encircled area in panel (a). Panel (c) shows the exciton-impurity 2D scattering cross sections for exciton $\gamma$ in Figure (2a) with $|\k|a = 4\times10^{-5}$ (solid line), $|\k|a =4\times10^{-3}$ (dashed line), and $|\k|a =4\times10^{-2}$ (dotted line). The calculations are for an array of LiCs molecules with one LiRb impurity. The lattice constant is $a = 400$ nm and $\mathcal{E}_0 = 3228.663$ V/cm.}
\label{fig:cross section}
\end{figure}

If a small number of LiCs molecules are replaced with molecules of a different kind, such as LiRb, the translational symmetry of the lattice is disturbed, as in a solid crystal with impurities (see Figure 1). Let $\Delta E_{eg}$ be the rotational excitation energy in host molecules, $\Delta E_{eg} + V_0$ the energy of the same transition in impurities, and $\Delta J_{n,m}$ the difference between impurity-host and host-host coupling constants for molecules at sites $n$ and $m$. Consider first the lattice with a single impurity at site $n=0$, which is described by the Hamiltonian \cite{Dubovskii:1965}
\beq
 \hat H =\hat H_0 + V_0\creation{B}{0}\annihilation{B}{0}+\sum_{n\neq0}\Delta J_{n,0}\left(\creation{B}{n}\annihilation{B}{0}+\annihilation{B}{n}\creation{B}{0}\right),
 \label{single impurity hamiltonian}
 \eeq
where $\hat H_0$ is the free-exciton Hamiltonian and $\creation{B}{0}$ and $\annihilation{B}{0}$ are the exciton creation and annihilation operators defined in the Methods section below. The exciton-impurity interaction can be described as a $\delta$-function potential with strength $V_0$, and a perturbation due to the difference in the dipole moments of host and impurity molecules.


We propose to modify $V_0$ by shifting the rotational levels of host and impurity molecules simultaneously using a static electric field.  Figure \ref{fig:cross section}a shows the excitation energies $\Delta E_{eg}$ for the rotational transition $N=0\rightarrow1$ of CsF, LiCs and LiRb molecules as functions of the electric field and demonstrates that the excitation energies for these combinations of molecules become the same at certain values of the electric field. 
 The interaction of polar molecules with the electric field allows for the possibility to explore the dependence of the exciton-impurity scattering cross section not only on the exciton wavevector $k$, but also on the scattering strength $V_0$, which is not possible in conventional solids. Figure \ref{fig:cross section}c shows the cross section for elastic scattering of an exciton by a single impurity in a 2D crystal as a function of the electric field, demonstrating resonant enhancement of the scattering cross section occurring for values of $V_0$ that support a shallow bound state (see Methods). These resonances are analogous to Feshbach resonances in atomic collisions, commonly used to tune the scattering properties of ultracold atoms \cite{feshbach}. Figure 3c shows that the scattering properties of excitons in the system proposed here can be tuned by varying the external electric field. 

Consider now a lattice with multiple impurities. It is well known that quantum particles in the presence of a random distribution of scattering centers undergo coherent localization \cite{localization}. In the system proposed here, the localization of excitons can be tuned by an electric field. As an example, we consider  localization of excitons in a 1D array of LiCs separated by 400 nm, with a random homogeneous distribution of LiRb impurities (see Figure 4). The Hamiltonian describing an exciton in the presence of $N_i$ substitutional impurities at positions $\vec{i}_n$ can be written as $\hat H = \hat H_0 + \hat W + \hat V$, with the corresponding matrix elements in the basis of free-exciton states $\ket{\Psi_{\k}}$ given by 
\begin{subequations}
\beq
\langle\hat H_0\rangle_{\q,\k}=E(\k) \delta_{\k,\q},\\
\eeq
\beq
\langle\hat W\rangle_{\q,\k} = \frac{2\Delta J(a)}{N_{\text{mol}}}\left(\cos{\q\cdot\vec{a}}+\cos{\k\cdot\vec{a}}\right) \sum_{\vec{i}_n=1}^{N_i} e^{i(\q-\k)\cdot \vec{i}_n},
\eeq
\beq
\langle\hat V\rangle_{\q,\k}=\frac{V_0}{N_{\text{mol}}}\sum_{\vec{i}_n=1}^{N_i} e^{i(\q-\k)\cdot \vec{i}_n},
\eeq
\label{many impurities hamiltonian}
\end{subequations}
where $E({\k})$ is the energy of the exciton, $\k$ and $\q$ denote exciton wavevectors and $\Delta J(a)=\Delta J_{n,n-1}$. If the impurities are distributed randomly, the terms $\hat V$ and $\hat W$ correspond to diagonal and off-diagonal disorder, respectively.
\begin{figure}[ht]
\includegraphics[width=0.47\textwidth]{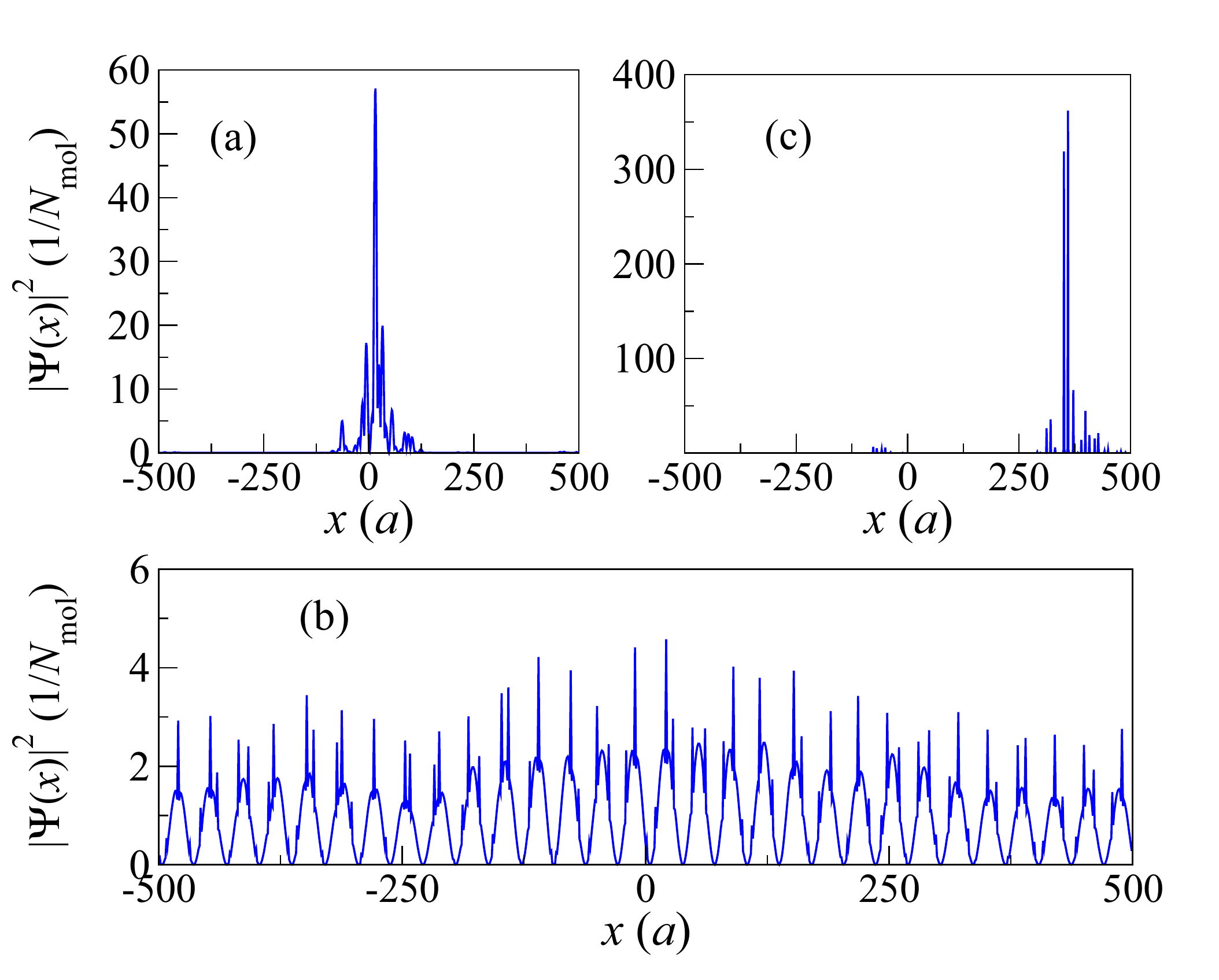}
\caption{Probability density $|\Psi(x)|^2$ describing an exciton near the top of the energy spectrum for a 1D array of 1000 LiCs molecules with 10\% of homogeneously and randomly distributed LiRb impurities. Panels correspond to different values of $V_0$: (a) $V_0= 0$, (b) $V_0/h= 22$ kHz, and (c) $V_0/h= 100$ kHz.
Tuning $V_0$ by an electric field can thus be used to realize different regimes of exciton dynamics from free propagation to strong localization in different spatial arrangements. The difference of the dipole moments of LiCs and LiRb molecules leads to the value $\Delta J = - 6.89$ kHz. 
}
\label{fig:time independent}
\end{figure}

 When an external electric field is such that $V_0= 0$, exciton-impurity scattering occurs only due to the difference in dipole moments between host and impurity molecules. Figure \ref{fig:time independent} shows the probability density of a particular eigenstate of Hamiltonian (\ref{many impurities hamiltonian}) near the top of the energy spectrum, for different values of $V_0$. Due to the negative effective mass of free-exciton states, high energy eigenstates are dominated by free-exciton states with $\k\approx 0$. These eigenstates are localized (Fig. \ref{fig:time independent}a). Delocalization of these states can be achieved by applying an electric field so that $V_0 \approx -4 \Delta J(a)$ (see Fig. \ref{fig:time independent}b and Eqs. (\ref{many impurities hamiltonian})). In this case, for a given $\k$, the matrix elements $\langle\hat V\rangle_{\q,\k}$ and $\langle\hat W\rangle_{\q,\k}$ cancel for $\q\approx\k$, which suppresses the coupling between the corresponding free-exciton states.
 Localized eigenstates with different energies become delocalized at different values of $V_0$. The wavepackets are localized for values of $V_0$ that do not balance the effect of $\Delta J(a)$ (Fig. \ref{fig:time independent}c). Figure (\ref{fig:time independent}b) shows an example of delocalization in a 1D disordered system due to correlations between diagonal and off-diagonal disorder \cite{Tessieri:2001}. 
 \\

\noindent
{\bf Applications}

The system proposed here offers three unique features: (i) long-lived excitons that are stable against spontaneous decay and whose effective mass can be controlled by an external electric field; (ii) dynamically tunable impurities; (iii) possibility to arrange impurities and host molecules in various configurations and dimensions. This opens up new possibilities for quantum simulation of fundamental physical phenomena:

Exciton - impurity scattering leads to localization of excitons in crystalline solids \cite{localization}. 
The exciton - impurity interactions in the system proposed here can be tuned from resonantly enhanced (Figure 3c) to entirely suppressed (Figure 4b) scattering, which can be used to study the dynamics of exciton localization, including the timescales for the formation of localized states and their dependence on exciton bandwidth and exciton - impurity interaction strength.

The negative effective mass of the excitons may be used for studies of negative refraction of microwave fields. At present, the materials exhibiting negative refraction are artificial metamaterials \cite{metamaterials}, in which the possibility to use a macroscopic (spatially averaged) description of electromagnetic fields is questionable \cite{agranovich2}. We also note that  the localization of quasi-particles with negative effective mass, which has not yet been observed, may exhibit different mechanisms of quantum interference due to multiple scattering events. Measuring the dynamics of excitons with tunable effective mass may provide new insights into the mechanism of Anderson localization \cite{localization}.

Controlled spatial distributions of impurities and molecular crystals with specific arrangements of crystal particles in one, two or three dimensions may be used to study the effects of dimensionality and finite size on energy transfer in mesoscopic materials \cite{Scholes:2006}. In addition, the system proposed here is ideally suited for the study of long- and short-range correlations in the disordered potentials. 
The presence of short- and long-range correlations in the disordered potentials may result in the appearance of a discrete
\cite{Dunlap:1990, Phillips:1991} or even continuous \cite{Izrailev:1999, Kuhl:2000} set of delocalized states in low-dimensional crystals.  Measurements of exciton localization in a crystal with tunable impurities can be used for time-domain quantum simulation of disorder-induced localization and delocalization of quantum particles.


The localized states displayed in Figure 4 are many-body entangled states of the molecules in an optical lattice. The possibility to tune exciton - impurity interactions can thus be exploited for the controlled creation of many-body entangled states of ultracold molecules, necessary for the experimental realization of quantum computation with molecular ensembles \cite{demille}.

Finally, a mixture of ultracold molecules with impurities forming a sublattice may be used to study the formation of wavevector space crystals of excitons. The eigenstates of such a two-species lattice correspond to a discrete distribution in $\k$-space. Tuning the impurities by a sinusoidally varying electric field could then be used to induce resonant transitions between different $k$-states. In particular, it could be interesting to couple reversibly the low-$k$ optically active, but slow (small group velocity) states with optically inactive, but fast (larger group velocity) states from the middle of the Brillouin zone. This would lead to the creation of excitonic wave packets, which can be slowed down or accelerated by an external electric field. \\

\noindent
{\bf Methods}

\noindent
{\it Dispersion curves} \\
The Hamiltonian describing the optical lattice with identical molecules in the presence of a static electric field is
\beq
\hat{H}=\sum_{n=1}^{N_{\text{mol}}} \left(B_e\hat{N}_n^2-\vec{d}_n\cdot\vec{E}\right)+\frac{1}{2}\sum_{n=1}^{N_{\text{mol}}}\sum_{m\neq n}^{N_{\text{mol}}}\hat{V}_{dd}(\vec{r}_n-\vec{r}_m),
\label{lattice hamiltonian}
\eeq
where $\vec{r}_n$ is the position of the {\it n}-th lattice site, $\hat{N}_n$ is the rotational angular momentum of the molecule, $B_e$ is the rotational constant, $\vec{d}_n$ is the
electric dipole operator, $\vec{E}$ is a static electric field, $\hat{V}_{dd}$ is the dipole-dipole interaction between molecules in different lattice sites, and $N_{\text{mol}}$ is the
total number of molecules. Hamiltonian (\ref{lattice hamiltonian}) can be written as \cite{Agranovich:2008}
\begin{eqnarray}
\hat H &=& \Delta E_{eg} \sum_{n} \crea{B}{n}{M}\anni{B}{n}{M} + \frac{1}{2}\sum_{n, m\neq n}J^{M,M}_{n,m}\crea{B}{n}{M}\anni{B}{m}{M}\nonumber\\
&&+ \frac{1}{2}\sum_{M,\;M'\neq M}\sum_{n, m\neq n}J^{M,M'}_{n,m}\crea{B}{n}{M}\anni{B}{m}{M'}.
\label{exciton hamiltonian}
\end{eqnarray}
The exciton creation operators are defined by $\crea{B}{n}{M}\ket{g}_n=\ket{e,M}_n$, where $\ket{g}$ and $\ket{e,M}$ denote ground and excited field-dressed rotational states, which
have a well-defined projection $M$ of the rotational angular momentum $\hat N$ along the direction of $\vec{E}$. The excitation transfer between molecules at lattice sites $\vec{r}_n$ and
$\vec{r}_m$ is described by the coupling constant $J^{M,M'}_{n,m}=\bra{e,M}_n\bra{g}_m \hat{V}_{dd}\ket{g}_n\ket{e',M'}_m$. In Eq. (\ref{exciton hamiltonian}), we have neglected the gas - condensed matter shift in the diagonal terms since it is much smaller than $\Delta E_{eg}$. 

If $\vec{E}$ is perpendicular to the intermolecular axis $\vec{R} = \vec{r}_n - \vec{r}_m$, then the dipole-dipole interaction between molecules $n$ and $m$ couples two-molecule states with
$\Delta(M_{n}+M_{m})=0,\pm2$, and if $\vec{E}$ is parallel to $\vec{R}$, the selection rule is $\Delta(M_{n}+M_{m})=0$.

By using $\crea{B}{n}{M} = \frac{1}{\sqrt{N_{\text{mol}}}}\sum_{\k}\crea{B}{\k}{M}e^{-i\k\cdot\vec{r}_n}$, where $\k$ is the exciton wavevector, Hamiltonian (\ref{exciton hamiltonian}) can
be rewritten as
\beq
\hat {H}= \sum_{\k}\sum_{M,M'}\left[\Delta E_{eg}\delta_{M,M'} + L_{M,M'}(\k)\right]\crea{B}{\k}{M}\anni{B}{\k}{M'},
\label{exciton hamiltonian 2}
\eeq
with $L_{M,M'}(\k)=\sum_n J^{M,M'}_{n,0 }e^{i\k\cdot\vec{r}_n}$.
Hamiltonian (\ref{exciton hamiltonian 2}) is diagonalized by the unitary transformation $\crea{B}{\k}{M} = \sum_{\mu}\alpha_{M,\mu}\crea{A}{\k}{\mu}$ \cite{Agranovich:2008} . The resulting
Hamiltonian is $\hat{H} = \sum_{\k,\mu}E_{\mu}(\k)\crea{A}{\k}{\mu}\anni{A}{\k}{\mu}$, where $\mu=\alpha, \beta, \gamma$ labels the dispersion relations associated with the eigenstates.

In the case where $\vec{E}$ is parallel to the molecular array, the eigenstates of Hamiltonian (\ref{exciton hamiltonian 2}) are: $\ket{\alpha_{\k}}=\ket{\Psi_{\k,M=1}}$,
$\ket{\beta_{\k}}=\ket{\Psi_{\k,M=-1}}$, and $\ket{\gamma_{\k}}=\ket{\Psi_{\k,M=0}}$. If the electric field $\vec{E}$ is perpendicular to the molecular array, the eigenstates are: $
\ket{\alpha_{\k}}=\tfrac{1}{\sqrt{2}}\left[\ket{\Psi_{\k,M=1}}-\ket{\Psi_{\k,M=-1}}\right]$, $ \ket{\beta_{\k}}=\tfrac{1}{\sqrt{2}}\left[\ket{\Psi_{\k,M=1}}+\ket{\Psi_{\k,M=-1}}\right]$, and
$\ket{\gamma_{\k}}=\ket{\Psi_{\k,M=0}}$.
The dispersion curves $E_{\mu}(\k)$ of the three exciton branches are shown in Fig. \ref{fig:dispersion curves} for a 1D array of LiCs molecules ($d_0 = 5.529$ Debye, $2B_e = 11.7998$ GHz
\cite{Aymar:2005}) with lattice constant $a= 400$ nm in the presence of an electric field parallel and perpendicular to the array. Analogous dispersion curves can be obtained in 2D and 3D and for other directions of the electric field. 
\\

\noindent
{\it Scattering resonances of excitons} \\
Figure \ref{fig:cross section}c shows the 2D exciton-impurity elastic cross section as a function of the electric field, obtained in the effective mass approximation $E(k) = E(0) - T_{k}$, for negative effective mass. $T_k = \hbar^2 |\k|^2 / 2|m_*|$ is the kinetic energy of the exciton. 
The impurities are LiRb molecules with  $d_0 = 4.165$ Debye, $2B_e = 13.2268$ GHz \cite{Aymar:2005}. If $ E_b$ is the energy of the bound state produced by $V_0$, and $\Delta$ is the exciton bandwidth, we write the scattering cross sections for $|\k| \ll \pi/a$, in 3D and 2D, as
\begin{subequations}
\beq
\sigma_{3\text{D}}(k,V_0)= \frac{2\pi \hbar^2/|m_*|}{T_k +  E_b^{(3\text{D})}},
\eeq
\beq
\sigma_{2\text{D}} (k,V_0)= \frac{4\pi^2/k}{\pi^2 +  \ln^2 \left[ \frac{E_b^{(2\text{D})} (\Delta-T_k)}{ T_k(\Delta- E_b^{(2\text{D})})}\right]},
\eeq
\label{cross sections}
\end{subequations}
where $ E_b^{(3\text{D})} = (2/\pi - V_0/\Delta)^2 \Delta$ and $E_b^{(2\text{D})} = \Delta/\left[ \exp(4\Delta /\pi V_0)-1 \right]$. In 3D, the bound state exists only if $V_0 > 2\Delta/\pi$. 
In 2D, any positive $V_0 < 0.2 \Delta$ supports an exponentially shallow bound state.  Resonant enhancement of the scattering cross section occurs for values of $V_0$ that support a shallow bound state ($E_b\sim T\rightarrow 0$). In 1D, shallow bound states appear only for vanishingly small $V_0$ and resonances may not be observable. As a result of the negative effective mass of the exciton, the bound state that leads to resonance is produced by a repulsive potential.  
Equations (\ref{cross sections}) are derived in the approximation $\Delta J_{n,0}=0$.  
Including $\Delta J_{n,0}$ in the calculation leads to a shift of the positions of the resonance \cite{Dubovskii:1965} and the resonant enhancement of the scattering cross section at a slightly different value of $V_0$. 
\\

\noindent
{\bf Acknowledgments} \\
We thank Evgeny Shapiro for discussions. This work is supported by NSERC of Canada and the Peter Wall Institute for Advanced Studies. 


\end{document}